\documentclass[a4paper]{jacow}

\ifboolexpr{bool{xetex} or bool{luatex}}
 {}
 {\usepackage[utf8]{inputenc}}

\usepackage{subfig}
\usepackage{graphics}
\usepackage{amssymb}
 \usepackage{graphicx}
\usepackage{marginnote}
\usepackage{float}
\usepackage{lipsum}

\ifboolexpr{bool{jacowbiblatex}}
 {%
  \addbibresource{jacow-test.bib}
  \addbibresource{biblatex-examples.bib}
 }{}

\copyrightspace[2cm] 

\begin{document}

\title{Modeling of an Electron Injector for the AWAKE Project}

\author{O. Mete\thanks{oznur.mete@manchester.ac.uk }, G. Xia,
\thanks{and The Cockcroft Institute, Sci-Tech Daresbury, Warrington, UK}The University of Manchester, Manchester, UK \\ R. Apsimon, G. Burt, $^{\dagger}$Lancaster University, Lancaster, UK \\ S. Doebert, CERN, Geneva, Switzerland, \\R. Fiorito, C. Welsch, $^{\dagger}$The University of Liverpool, Liverpool, UK}

\maketitle

\begin{abstract}
Particle-in-cell simulations were performed by using PARMELA to characterise an electron injector with a booster linac for the AWAKE project in order to provide the baseline specifications required by the plasma wakefield experiments. Tolerances and errors were investigated. A $3\,$GHz travelling wave structure designed by using CST code. Particles were tracked by using the field maps acquired from these electromagnetic simulations. These results are presented in comparison with the generic accelerating structure model within PARMELA. 
\end{abstract}

\section{Introduction} 
The AWAKE project is a proton driven plasma wakefield acceleration experiment by utilising the driver proton beam from CERN's SPS injector and a custom photo injector for the witness (trailing) electron beam \cite{awake_cdr, ipac2014}.

\begin{figure*}[htb!] 
\centering
\includegraphics[width=0.6\textwidth] {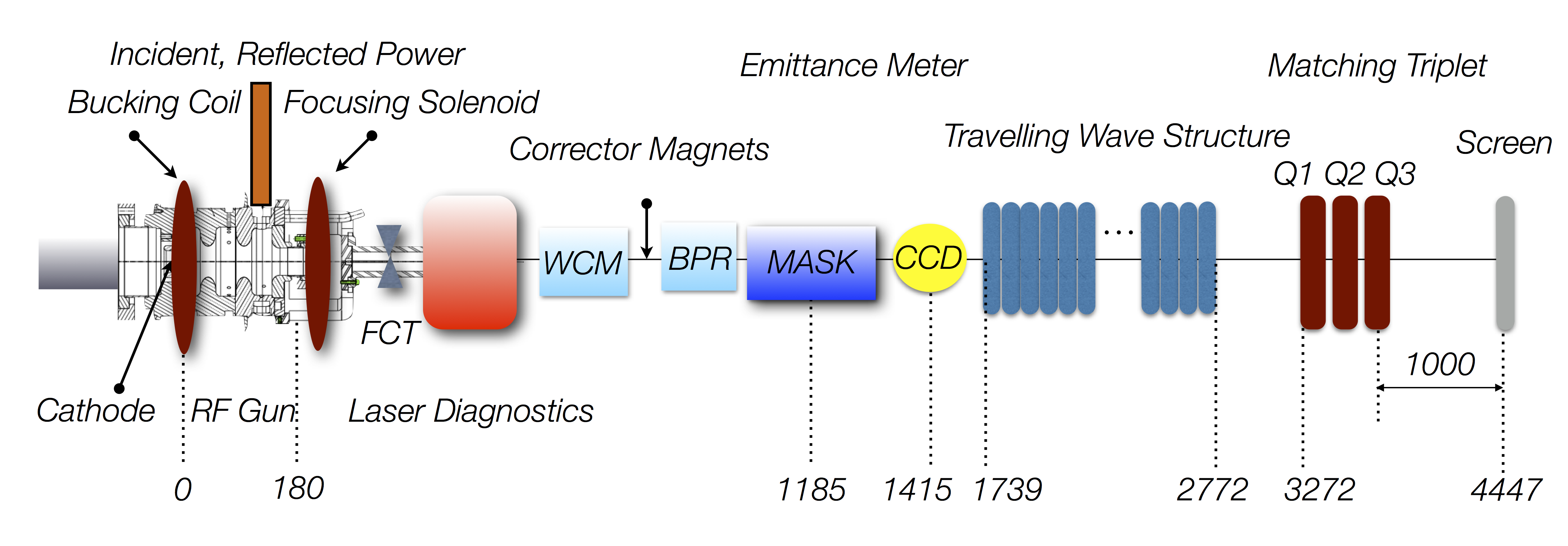}
\caption{The layout of the photo injector and the booster linac section to provide a witness beam for the AWAKE plasma wakefield acceleration experiment.}
\label{fig:layout}
\vspace{-1.5em}
\end{figure*}

Figure \ref{fig:layout} shows the layout of the injector consisting of an S-band standing wave RF gun (SW), previously used in CERN's PHIN photo injector \cite{phingun}, followed by laser optics to direct the laser onto the cathode. Beamline continues with beam current and position monitors and a pepper pot emittance measurement system suitable for a beam subject to space charge effects. A new S-band booster section (accelerating travelling wave structure, ATS) was designed by using CST suite \cite{CST} and introduced in the model to boost the beam energy to a region adjustable between $16-20\,$MeV. ATS is followed by a quadrupole triplet and a downstream screen to perform quadrupole scans for emittance measurement.

The tracking studies have been performed by using PARMELA \cite{parmela_manual}. For ATS the model provided by PARMELA and field maps extracted from CST were used and compared. Implementation of CST maps into PARMELA are discussed in the following sections.
\section{Booster Linac} 
\begin{figure}[htb!] 
\centering 
\includegraphics[width=0.35\textwidth] {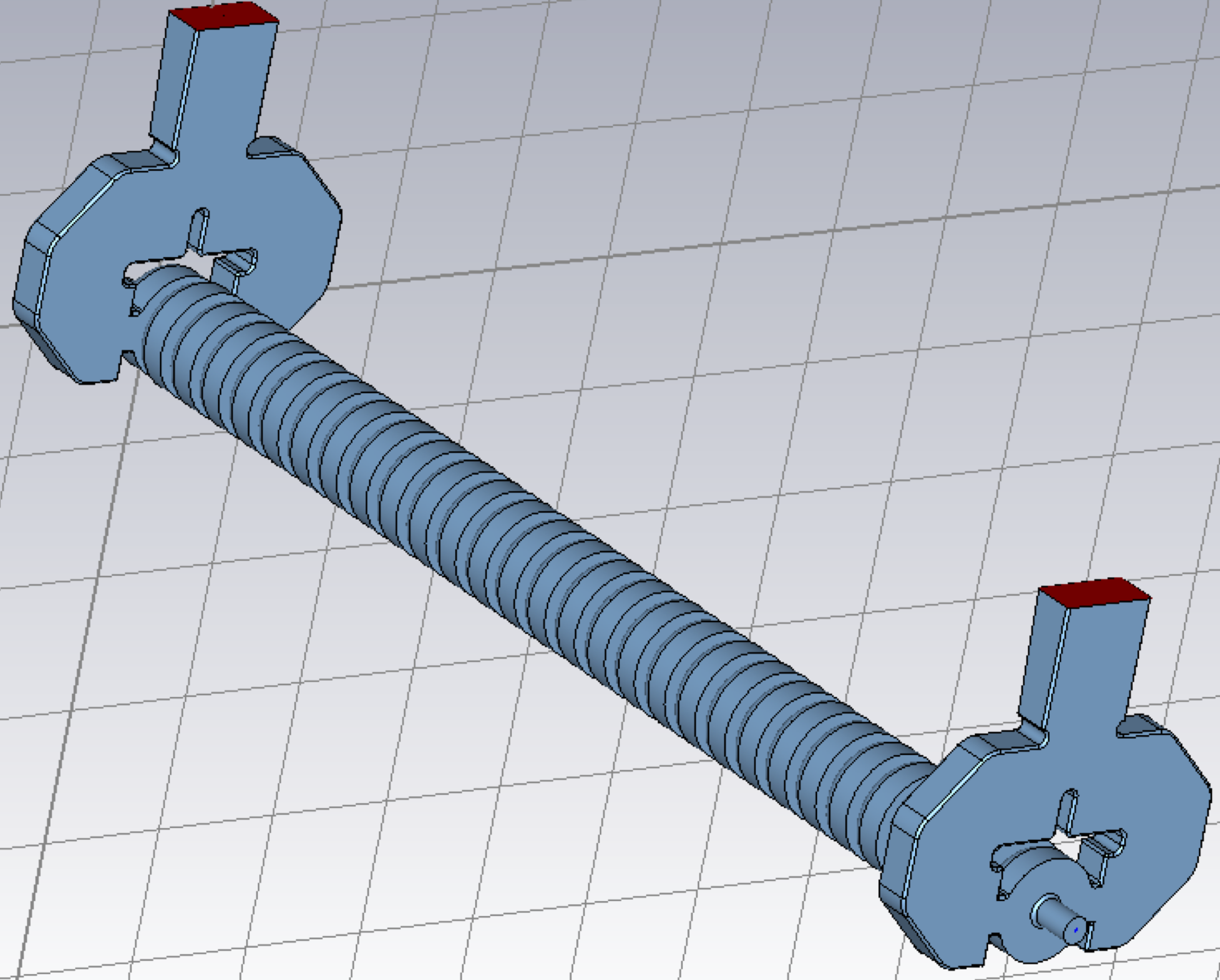}
\caption{A snapshot from CST design of the S-band travelling wave structure as the energy booster linac for the electron injector.}
\label{fig:ATS}
\vspace{-1.5em}
\end{figure}

An S-band booster linac, ATS, was designed as a travelling wave structure with constant gradient of $15\,$MV/m through the entire structure (Fig.\ref{fig:ATS}) . It consists of $30\,$cells with $120^{\circ}$ phase advance and varying radii matched to $1\,\mu$m precision. ATS was optimised for low reflection coefficient of about $2.5\%$. The multipole terms \cite{Olave} due to transverse RF-kicks are $ 9.4241\times10^{-7}\,$mT, $7.8418\times10^{-5}\,$mT/m and $4.9\times10^{-3}\,$mT/m$^2$, respectively, from dipole to sextuple terms. 
\section{Using CST Field Maps in PARMELA} 
In order to use CST field maps in PARMELA tracking simulations, a MATLAB \cite{matlab} script was prepared to format the standard CST field maps into the form required by PARMELA . The information that PARMELA requires on each line of a field map can be found in Table IV-2 from the program manual \cite{parmela_manual}. 

In PARMELA, two field maps must be provided for a travelling wave structure; one produced with Neumann boundary condition (cosine map) and the other with Dirichlet boundary condition (sine map). These fields which are shifted in phase by $90^{\circ}$ are fed into PARMELA by using the TRWCFIELD command. A single TRWAVE line is used to represent the entire ATS including the bore tubes with lengths equal to a cell length at each end of ATS to account for the fringe fields.
\begin{figure}[htb!] 
\centering
\subfloat[]{\includegraphics[width=0.45\textwidth] {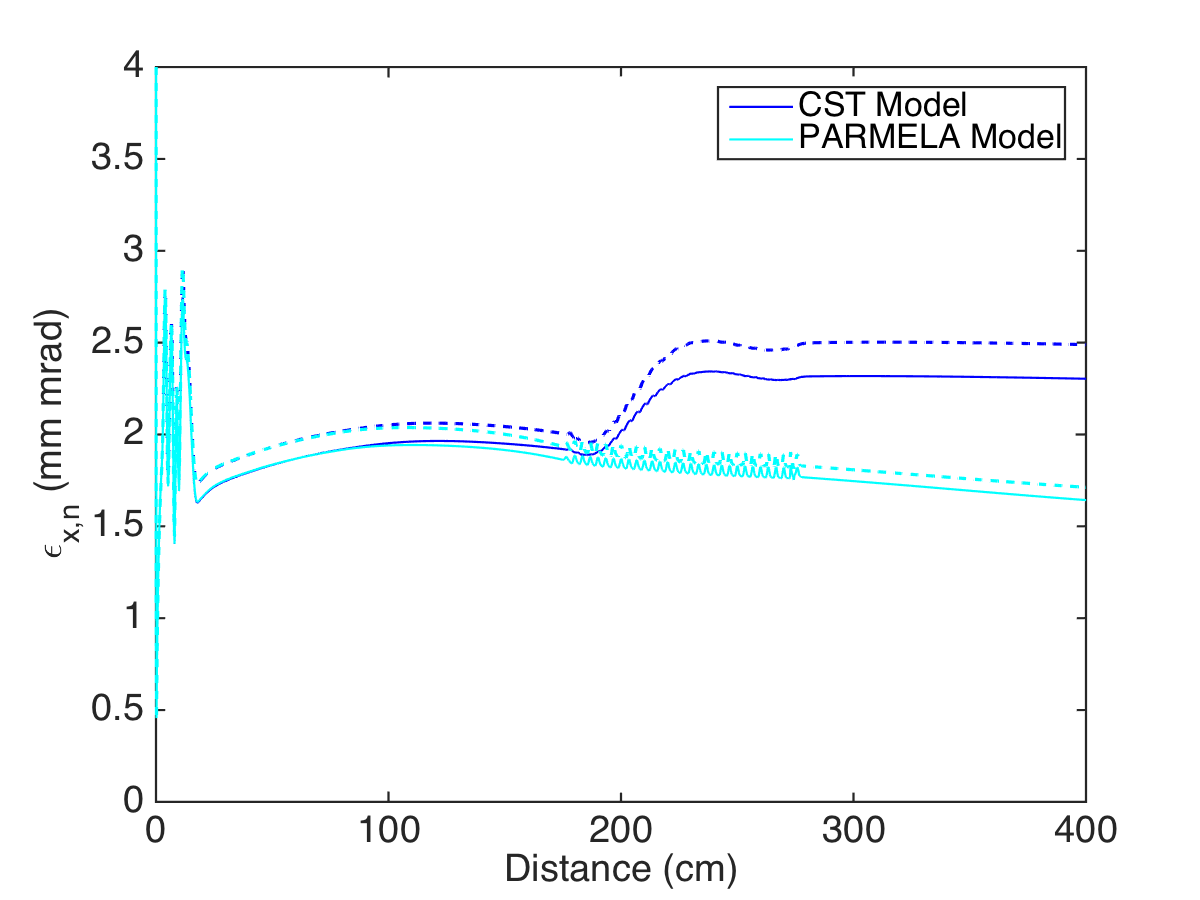}} \\
\subfloat[]{\includegraphics[width=0.45\textwidth] {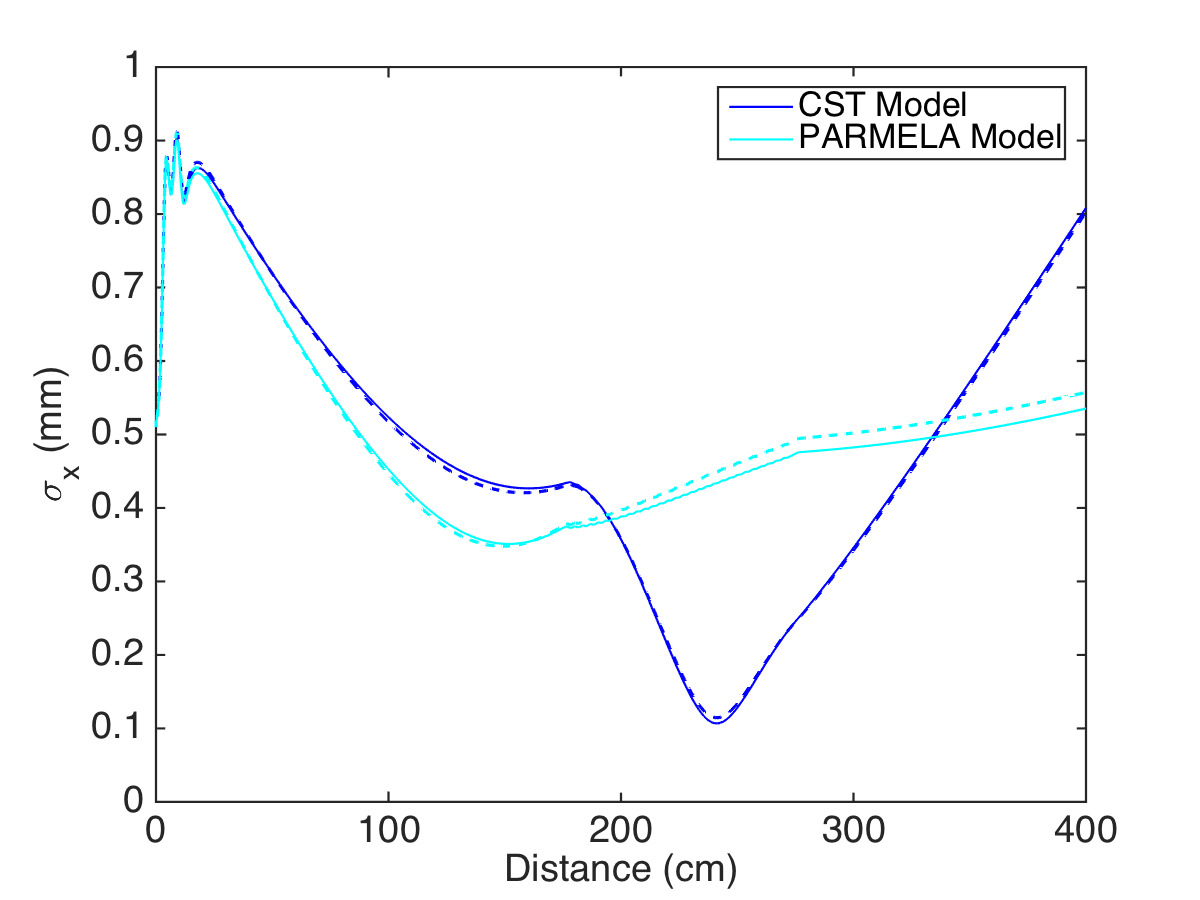}} 
\caption{a) Transverse normalised emittance and b) the beam size as a function of the distance from the cathode. (Solid and dashed lines are x and y axes, respectively.)}
\label{fig:trans_dyn}
\vspace{-1.5em}
\end{figure}
\begin{figure}[htb!] 
\centering
\includegraphics[width=0.45\textwidth] {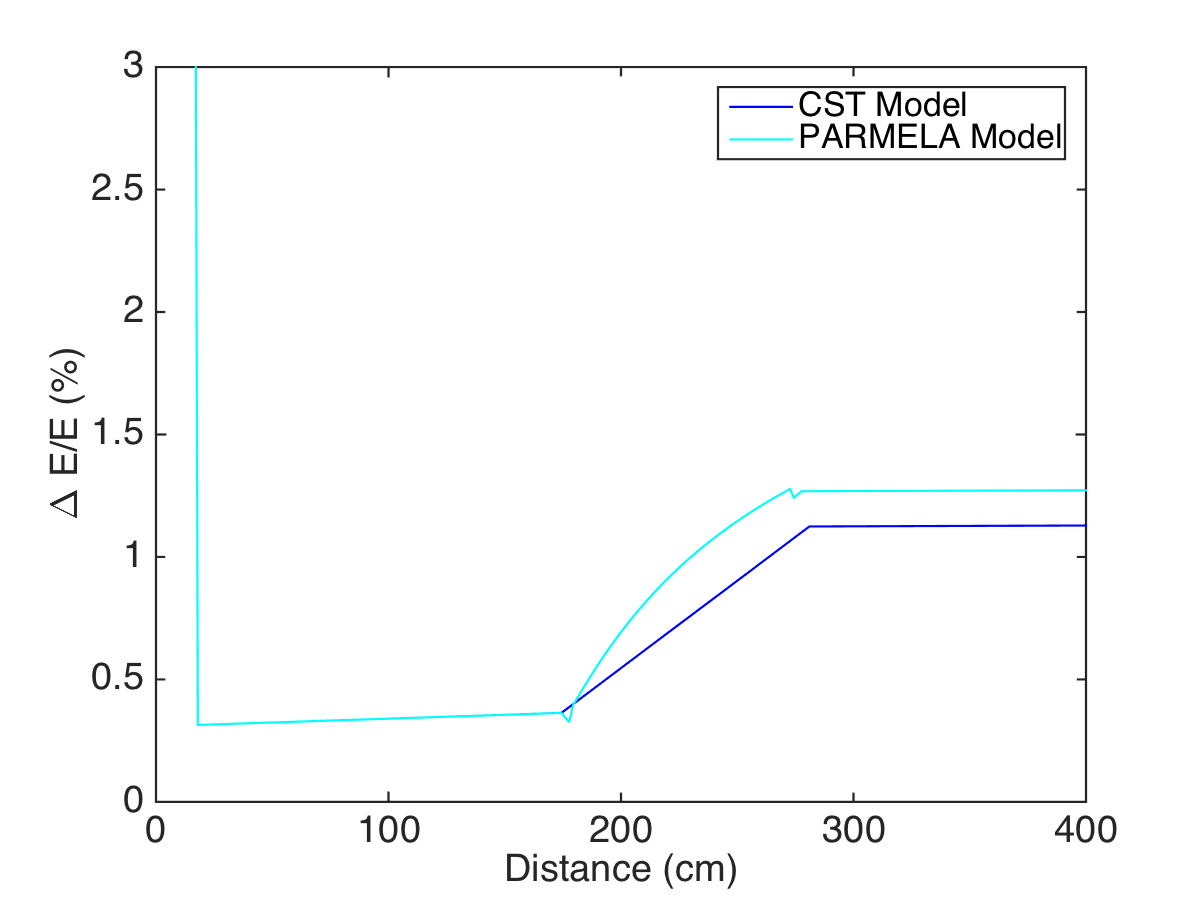}
\caption{Evolution of the beam energy spread as a function of the distance from the cathode.}
\label{fig:long_dyn}
\end{figure}
\section{Baseline Design} 
In order to maintain the balance between the emittance growth and the energy spread within ATS, a $10^{\circ}$ off-crest phase was chosen for the baseline design by using the CST maps of the current design while the SW is optimised for the highest energy. In addition, the same beam dynamics conditions can be met with on-crest RF phase in the case of the PARMELA model for a travelling wave structure. 

The space charge component of the emittance is compensated using the field produced by two solenoids located around the SW. Slightly different working points to satisfy the constant envelope condition were determined for the field distributions from PARMELA and CST models. 

The beam dynamics specifications are presented in Table \ref{tbl:parameters} in comparison with these two.
\begin{table}[hbt!]
\centering
\caption{Required specification and values produced with different models.}
	\resizebox{8cm}{!}{ 
	\begin{tabular}{lcccc}
	\toprule 
	Parameter  &  Required  & PARMELA Map  & CST Map\\
	\midrule
	E (MeV)                 			&16  	 &  16             & 16\\
	$\Delta E$/E ($\%$)  		& 0.5	 &  1.3	     & 1.1\\
	$\epsilon_{x,y}$ (mm mrad)      & 2	 &   1.7,1.8    &  2.3, 2.5\\
	$\beta_{x,y}$ (m)         	        & 5	 &  4, 4	     & 0.8, 0.8\\
	$\alpha_{x,y}$			        & 0	 &  -0.2,-0.2   & -1.4, -1.2\\   
	\bottomrule
	\end{tabular}}
\label{tbl:parameters}
\end{table}
\section{RF Phase Errors} 
Variations in the beam dynamics due to the RF phase errors induced on SW and ATS were investigated. As they are planned to be fed by the same klystron, the same phase error can be assigned to both structures. A typical phase error of $300\,$fs ($1\sigma$) was induced over a $500$ samples assuming the RF jitter as the error source. Resulting variations were presented in Table \ref{tbl:phase_errors}. Especially time of flight errors are crucial to determine as they affect the synchronisation in the entrance of the plasma cell. 
\begin{table}[hbt!]
\centering
\caption{Implications of the phase errors on various beam dynamics observable.}
	\resizebox{8cm}{!}{ 
	\begin{tabular}{lcccc}
	\toprule 
	Parameter  &  $1\sigma$ Error  & Error/Degree of the phase jitter       \\
	\midrule
	E (keV)                 			&1.4  	 &  5            \\
	$\Delta E$/E  (ppm)		       & 28	 	 &  93	     \\
	$\epsilon_{x,n}$ (nm-rad)         & 2     	 &   7    \\
	Time of flight (fs)        	        & 57		 &  190	     \\
	\bottomrule
	\end{tabular}}
\label{tbl:phase_errors}
\end{table}
\section{Emittance Diagnostics under Space Charge} 

\begin{figure}[htb!] 
\centering
\includegraphics[width=0.4\textwidth] {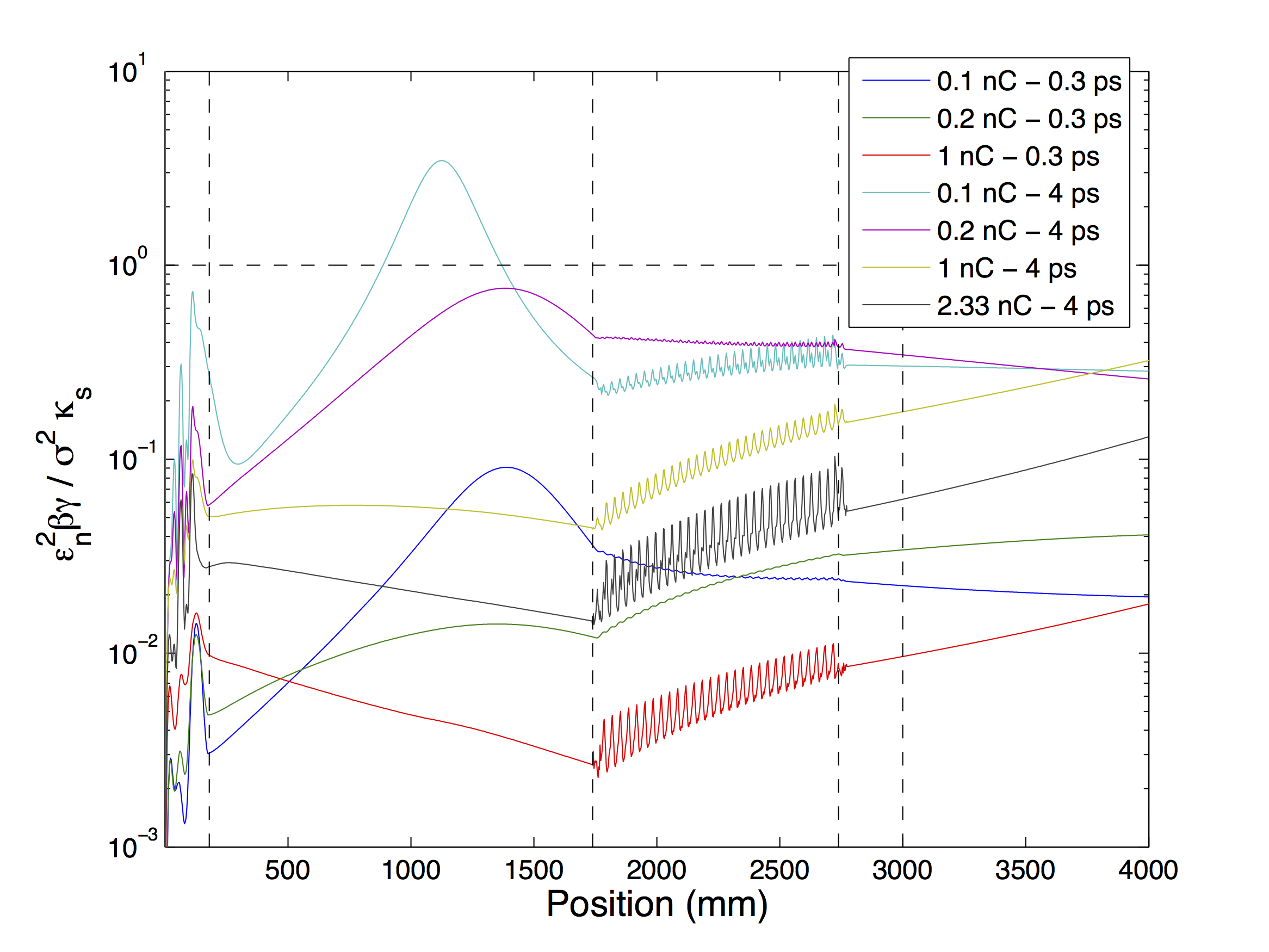}
\caption{The ratio of space charge and emittance term of the beam envelope as a function of the distance from the cathode.}
\label{fig:scheff}
\vspace{-1.0em}
\end{figure}

\begin{figure}[htb!] 
\centering
\includegraphics[width=0.35\textwidth] {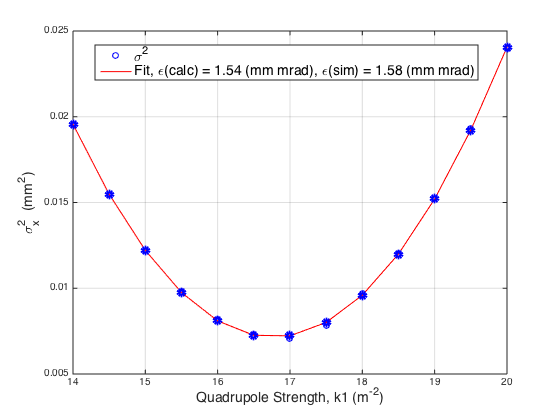}
\caption{A simulated quadrupole scan result in the post-booster region.}
\label{fig:quadscan}
\vspace{-1.0em}
\end{figure}

For a photo injector, the dominant limitation for the emittance diagnostics is the effect of space charge. Figure \ref{fig:scheff} shows the ratio of two defocusing terms from the envelope equation; the space charge and the outward pressure due to beam emittance. Different curves correspond to different bunch charges and lengths which are considered within the operational range of the injector. Although some cases are less space charge dominated ($>1$) than others, in all cases considered, space charge is the dominating component of the total beam emittance. 
In the pre-booster region, a pepper-pot measurement system will be implemented as a standard technique to measure the emittance of space charge dominated beams. It is currently being designed at the Cockcroft Institute. For the post-booster region quadrupole scan technique is planned. The feasibility and reliability of this technique was assessed by simulations. Figure \ref{fig:quadscan} compares the emittance values calculated from a simulated quadrupole scan and the rms emittance value at the same location.  The results differ $2\%$ implying that emittance can be reliably measured within measurement error by using quadrupole scan in the post-booster region.
\begin{figure}[htb!] 
\vspace{-1.5em}
\centering
\includegraphics[width=0.45\textwidth] {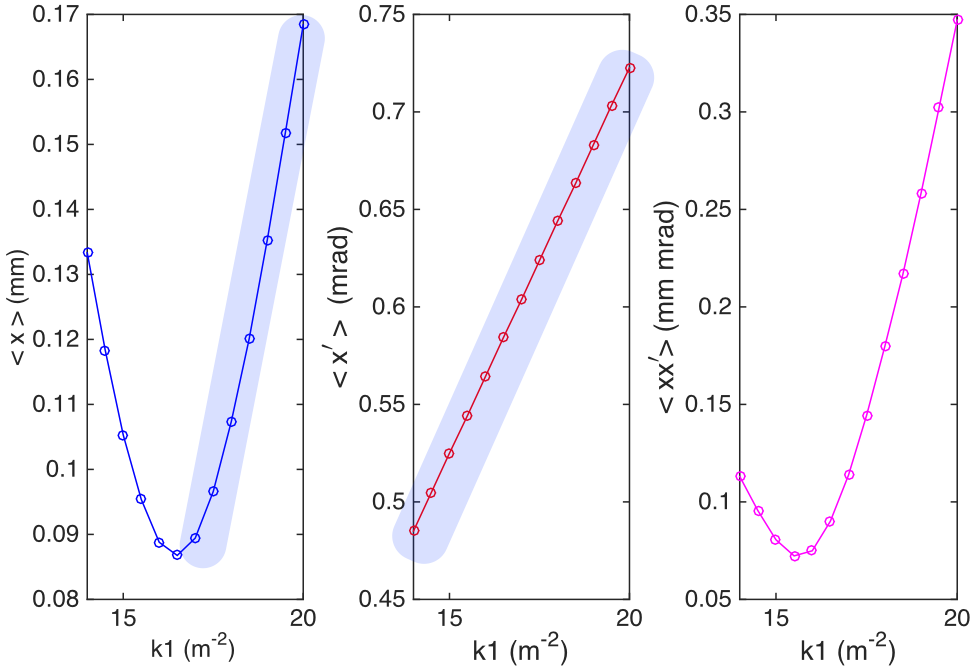}
\caption{Beam size, divergence and the correlation terms for the range of working points of the injector.}
\label{fig:divergence}
\vspace{-1.5em}
\end{figure}

We have also evaluated the Optical Diffraction Radiation Dielectric Foil Radiation Interference (ODR-DFRI) to measure the rms emittance. This technique which is a modified version of the Optical Transition Radiation Interference (OTRI) \cite{Fiorito1} technique extends the latter to lower beam energies and has the advantage of single shot operation. The technique can be used in conjunction with a new algorithm  developed to use sparse data from a quad scan to measure the rms emittance of space charge dominated beams \cite{Poorrezaie}. Thus, the lower limit of measurable divergence values is $0.3\,$mrad for the injector. Figure \ref{fig:divergence} shows the beam size, divergence and the correlations terms for a range of working points for the injector. As seen in the figure, the divergence values range from 0.5 and upwards implying that the ODR-DFRI technique could have been implemented for this system as a complementary technique to the conventional quadrupole scan in case of more severe space charge limitations.
\section{Conclusions and Outlook} 
The tracking simulations for the injector were performed by using the field models from both PARMELA and CST. Slight discrepancies in the results are under investigation. Errors due to phase jitter were studied. Emittance diagnostics under space charge effect is studied for both low and higher energies. 
\section{Acknowledgements}
This work was supported by the Cockcroft Institute Core Grant and STFC. Authors would like to express their gratitude to Janet Susan Smith and Stefano Mazzoni for useful communications on feasibility of beam diagnostics equipment during the course of this study. 
%
\raggedend

\end{document}